\documentclass[12pt]{article}

\input epsf
\usepackage{amssymb,amsmath,wrapfig,subfigure}
\usepackage[matrix,arrow,curve]{xy}
\usepackage{cite}
\usepackage{pdfsync}
\usepackage{graphicx,color}
\usepackage[debug,pageanchor=false]{hyperref}
\definecolor{link}{rgb}{.8,.15,.1}
\hypersetup{colorlinks=true,linkcolor=link,citecolor=link,urlcolor=link,linktocpage}

\makeatletter
\@addtoreset{equation}{section}
\makeatother

\def\rr {{\Bbb R}}

\def\del {\partial}

\def\del {\partial}

\def\be{\begin{equation}}
\def\ee{\end{equation}}
\def\bea{\begin{eqnarray}}
\def\eea{\end{eqnarray}}
\def\de{\partial}

\newcommand{\noname}{\varrho}
\newcommand{\notsure}{\mu}

\newcommand{\im}{{\rm Im}\,}
\newcommand{\re}{{\rm Re}\,}

\begin{document}

	       \begin{titlepage}

	       \begin{center}

	       \vskip .3in \noindent

	       {\Large \bf{Supersymmetry in Lorentzian Curved Spaces \\\vspace{.2cm} and Holography}}


\vskip 5mm
	       \bigskip

		 Davide Cassani$^1$, Claudius Klare$^2$, Dario Martelli$^1$,\\ Alessandro Tomasiello$^2$ and Alberto Zaffaroni$^2$\\

	       \bigskip
		
\vskip 7mm

			{\small $^1$ Department of Mathematics, King's College London, \\
The Strand, London WC2R 2LS, United Kingdom\\
{\footnotesize{\tt davide.cassani, dario.martelli AT kcl.ac.uk}}
			\vspace{.1cm}
			
			$^2$ Dipartimento di Fisica, Universit\`a di Milano--Bicocca,\\
 I-20126 Milano, Italy\\
	       and\\
	       INFN, sezione di Milano--Bicocca,\\
	       I-20126 Milano, Italy}\\
	       
{\footnotesize{\tt claudius.klare, alessandro.tomasiello, alberto.zaffaroni AT mib.infn.it}}


	       \vskip .5in
	       {\bf Abstract }
	       \vskip 1mm

	       \end{center}

We consider superconformal and supersymmetric field theories on four-dimensional 
Lorentzian curved space-times, and their five-dimensional holographic duals. 
As in the Euclidean signature case, preserved supersymmetry for a superconformal theory is equivalent to the existence of a charged conformal Killing spinor. Differently from the Euclidean case, we show that the existence of such spinors is equivalent to the existence 
of a null conformal Killing vector. For a supersymmetric 
field  theory with an R-symmetry, this vector field is further restricted to be Killing.
We demonstrate how these results agree with the existing classification of supersymmetric solutions of minimal gauged supergravity in five dimensions.


	       \noindent

	       \vfill
	       \eject


	       \end{titlepage}

\tableofcontents

\newpage 

\section{Introduction} 
\label{sec:intro}

Space-time curvature can be very safely neglected in the study of the non-gravitational interactions in nature. From a theoretical point of view, however,
 one can expect to learn interesting lessons about quantum field theory by studying it in curved spacetimes, just like one does for example by varying 
the gauge group rank, the number and representation of fields, or sometimes even the space-time dimension. 
The study of supersymmetric field theories on curved spaces with Euclidean signature has recently found various 
applications \cite{pestun-S4,kapustin-willett-yaakov,jafferis-Z,hama-hosomichi-lee,hama-hosomichi-lee2,imamura-yokoyama}.

When the field theories admit a holographic dual description, it is natural to study 
 supergravity solutions comprising an asymptotically locally anti-de Sitter (AdS) space-time,
where curved backgrounds arise as geometric data on the conformal boundary.
 For example, round $d$-dimensional spheres arise simply as the boundary of Euclidean AdS space in $d+1$ dimensions.  If the field theory possesses an Abelian R-symmetry, then more complicated backgrounds may be
obtained turning on a background gauge field coupled to the R-symmetry current, and gravity duals may be constructed in the framework of 
gauged supergravity. Examples in four dimensions were presented in \cite{martelli-passias-sparks,martelli-sparks-nuts}
and involve three-dimensional Chern--Simons gauge theories in the background of certain squashed three-spheres, with non-trivial R-symmetry gauge field. 
Examples in five dimensions were constructed earlier in Lorentzian signature \cite{gauntlett-gutowski,behrndt-klemm,gauntlett-gutowski-suryanarayana}, 
and we will discuss them in this paper as an illustration of our general results. 
In one case, the Lorentzian solution can easily be Wick-rotated to Euclidean signature. 

In \cite{ktz,dumitrescu-festuccia-seiberg} it was shown that four-dimensional superconformal theories (and, more generally,  supersymmetric theories with an R-symmetry) 
remain supersymmetric in four Euclidean dimensions when the background space is \emph{complex}. 
From the point of view of rigid supersymmetry, the Lorentzian signature case has so far been less studied; one exception is anti-de Sitter, which has been considered for a long time, 
since e.g.~\cite{keck,zumino-AdS,ivanov-sorin-superfield}, and more recently in e.g.~\cite{adams-jockers-kumar-lapan,jia-sharpe} (which also contain a more complete list of references). 
In this paper we consider the same question as in \cite{ktz}, namely under what conditions a supersymmetric field theory can preserve any supersymmetry on a curved space, in Lorentzian signature.\footnote{One might also attack this question using superspace, as in \cite[Chap.~6]{buchbinder-kuzenko}.}
We begin by considering superconformal theories. As in \cite{ktz}, we find very generally that the boundary $M_4$ needs to admit a 
\emph{conformal Killing spinor} (CKS) $\epsilon$, possibly charged under a gauge field $A_\mu$. The smallest amount of supersymmetry corresponds to $\epsilon$ being chiral; since the 
equation is linear, whenever $\epsilon$ is a conformal Killing spinor $i \epsilon$ is one too. So the minimal amount of supersymmetry is two supercharges; we focus on this case. 
As we will show, the condition on the geometry of $M_4$ for this to happen is very different  from the Euclidean case.
 Namely, $M_4$ has a conformal Killing spinor if and only if it has a \emph{null conformal Killing vector} $z$. 
The gauge field $A_\mu$ can then be determined purely from data of the metric on $M_4$.

One can also study supersymmetric theories on curved spaces using the method proposed in \cite{festuccia-seiberg}. This consists in coupling the theory to supergravity,
and then freezing its fields to background values. For a superconformal theory, the appropriate gravity theory is conformal supergravity \cite{kaku-townsend-vannieuwenhuizen1,kaku-townsend-vannieuwenhuizen2,kaku-townsend-vannieuwenhuizen3,ferrara-zumino-conformal}; we will show that 
the result of this procedure is again that 
$M_4$ should admit a conformal Killing spinor. For a supersymmetric theory with an R-symmetry which is not superconformal, it is natural to use new minimal
 supergravity \cite{sohnius-west}, where the off-shell gravity multiplet contains $g_{\mu\nu}$ and two vectors $a_\mu$, $v_\mu$ (the former coupling to the R-symmetry current). 
For the theory obtained by this procedure to be supersymmetric on a curved $M_4$, one should then solve an equation for $\epsilon$ which is (locally)
 equivalent to the CKS equation,
with a suitable map of $a_\mu$, $v_\mu$ with $A_\mu$ and some data of the geometry. 
This map in general produces a $v_\mu$ which is complex, which in Lorentzian signature is not acceptable; imposing that it should be real turns out to require that the 
conformal Killing vector $z$ is now actually a \emph{Killing vector}. As we will see, this stronger condition arises automatically from the bulk perspective when a certain natural 
choice of coordinates is used. 

After having determined that supersymmetry leads  to  clear geometrical requirements, 
one naturally wonders how this is related to the geometry in the bulk. The geometry of supersymmetric solutions 
of (Lorentzian) five-dimensional minimal gauged supergravity was considered in \cite{gauntlett-gutowski}, and it is interesting to compare our result to their classification. 
Indeed, one of the conditions found in \cite{gauntlett-gutowski} in the bulk was the existence of a Killing vector $V$, which may be time-like or null.  We will show that this vector always becomes null at the boundary, and reduces to the conformal Killing vector $z$.  We will also check that the other conditions from the bulk become redundant at the boundary, in agreement with our results. 

The rest of the paper is organized as follows. In section \ref{sec:ads} we show that supersymmetric asymptotically locally AdS solutions 
in the bulk imply the existence of a charged conformal Killing spinor on the boundary $M_4$.
In section \ref{sec:cks}  we show that such a spinor can exist if and only if $M_4$ has a null conformal Killing vector, and thus that this is the condition 
for a superconformal theory on $M_4$ to preserve some supersymmetry. In section \ref{sec:nm} we extend our analysis to theories which are not necessarily superconformal, 
but simply supersymmetric with an R-symmetry; we show that the condition on $M_4$ is now that it admits a null Killing vector. In sections \ref{sec:bulk} 
and \ref{sec:ex} we compare our results on $M_4$ with the bulk analysis of supersymmetric solutions of gauged minimal supergravity performed in \cite{gauntlett-gutowski}, and find agreement. 

      
\section{Conformal Killing spinors from the bulk} 
\label{sec:ads}
\label{sub:cks}

In this section we discuss how supersymmetry in a holographic gravity set-up implies the existence of a conformal Killing spinor
on the boundary geometry. The analysis is similar to the one performed in Euclidean signature in \cite{ktz}.

We consider  minimal gauged supergravity in five dimensions. 
In Lorentzian  signature $(-,+,+,+,+)$, the bosonic part of action
is\footnote{We use notations adapted from \cite{gauntlett-gutowski}, to which we refer for details. To compare
 with \cite{gauntlett-gutowski}, one has to identify $\chi = 2\sqrt 3  \ell^{-1}$. Moreover, one needs to switch between 
mostly plus and mostly minus signature, which means flipping the sign of the metric and 
taking $\gamma^\alpha_{\rm here} \,=\, -i \,\gamma^\alpha_{\rm there}$.
Also, we denote with a hat  five-dimensional quantities that might be confused with four-dimensional ones. \label{footnoteconventions}}
\be 
S \, = \, \frac{1}{4\pi G}  \int \left( \Big( \frac{1}{4} \hat R+ \frac{3}{\ell^{2}}\,\Big)*1 - \frac{1}{2} \hat F\wedge *\hat F - \frac{2}{3\sqrt3}  \hat F\wedge \hat F \wedge \hat A\ \right) \, ,
\ee
where $\hat F = d\hat A$ and $\ell \neq 0$ is a real constant. 
We are interested in supersymmetric solutions of this theory,  which are asymptotically locally AdS (with radius $\ell$),
 and in particular we will assume the following asymptotic Fefferman--Graham form of the metric
\bea
\ell^{-2} d\hat s^2 & = &  \frac{dr^2}{r^2} + r^2\left(  g_{\mu\nu} (x) +  \mathcal{O}(r^{-1})  + \dots  \right) dx^\mu dx^\nu \ , 
\label{feffer}
\eea
where $\mu,\nu=0,\ldots,3$ are curved indices and $g_{\mu\nu}(x)$ is a four-dimensional 
metric of Lorentzian signature $(-,+,+,+)$. The vielbein takes the form 
\be
\hat e^a \,=\, \ell\, r \,  e^a + \mathcal{O}(1) \,, \qquad \hat e^5 \,=\, \frac{\ell dr}{r}\ ,
\ee
where $a=0,\ldots, 3$ are flat four-dimensional indices and $ e^a = e^a_\mu (x)dx^\mu$ is a vielbein for $ g_{\mu\nu}(x)$.
The associated spin connection is
\be
\hat \omega^{ab} =  \omega^{ab} + \mathcal O(r^{-1})\,, \qquad \hat \omega^{a5} = r e^a + \mathcal O(r^{-1})\ .
\ee
For the bulk gauge field we assume
\begin{equation}\label{AnsatzA}
		\hat A_{\mu}(x,r)  =  -\frac{\ell}{\sqrt 3} A_\mu(x) + {\cal O}(r^{-1}) \ ,  \qquad \hat A_r(x,r)  =  0 \ ,
\end{equation}
which are compatible with the equations of motion. It follows that $\hat F_{\mu\nu} = \mathcal O(1)$ and $\hat F_{\mu r} = \mathcal O(r^{-2})$.
The Killing spinor equation corresponding to a vanishing gravitino variation is
\be\label{KillingSpEq}
\left[ \hat \nabla_\alpha + \frac{i}{4\sqrt 3}\left( \gamma_{\alpha}{}^{\beta\gamma} - 4\delta^\beta_\alpha\gamma^\gamma\right) \hat F_{\beta\gamma} \right]\epsilon^I +  \frac{1}{2\ell} \epsilon^{IJ} \big( i \gamma_\alpha + 2\sqrt 3 \hat A_\alpha \big) \epsilon^J  \,=\, 0\ ,
\ee
where we are using flat $\alpha, \beta$ five-dimensional spacetime indices. Our conventions 
for the spinors, which  are symplectic-Majorana, can be found in Appendix \ref{SpinConventions}.

At leading order in the asymptotic expansion in $r$, the radial part of the Killing spinor equation \eqref{KillingSpEq} 
gives rise to
\bea \label{eq:delrepsI}
\partial_r\epsilon^I  +  \frac{i}{2r}  \gamma_5 \,\epsilon^{IJ} \epsilon^J  \,=\, 0\,,
\label{seccom}
\eea
where the index on $\gamma_5$ is flat.
Note that the contribution of the gauge field strength obtained from \eqref{AnsatzA} is sub-leading and therefore drops out. Eq. (\ref{eq:delrepsI}) implies
 that the two symplectic-Majorana spinors take the asymptotic form 
\begin{equation}\label{eq:expansion}
	\begin{split}
		\epsilon^1 \,& =\,  r^{1/2}\epsilon + r^{-1/2}\eta \,+ \,\mathcal O(r^{-3/2}) \ ,\\
		\epsilon^2\, & =\,   i \gamma_5 (r^{1/2} \epsilon -  r^{-1/2}\eta)\, +\, \mathcal O(r^{-3/2}) \ ,
	\end{split}
\end{equation}
where $\epsilon$ and $\eta$ are independent of $r$. 
Plugging these expressions back into the remaining components of \eqref{KillingSpEq}, one finds that at leading order the spinors obey the following 
equation
\be
\Big(  \nabla_\mu - i   A_\mu \gamma_5 \Big) \epsilon +   \gamma_\mu \gamma_5 \eta \; = \; 0\label{bdryEq} \,.
\ee
In the gamma matrix representation we adopted (see appendix \ref{SpinConventions}), the symplectic-Majorana condition in 
five dimensions implies that the four-dimensional spinors obey
$\epsilon^* = \epsilon$ and $\eta^* = -   \eta$.
We can also use $\gamma_5$ to define the chirality for the boundary spinors,
\bea
\epsilon = \epsilon_+ + \epsilon_-\,, \qquad \quad \eta = \eta_+ + \eta_-\,,
\eea
where $\gamma_5 \epsilon_\pm = \pm \epsilon_\pm $ and $\gamma_5 \eta_\pm = \pm \eta_\pm $.
Taking the the trace of  \eqref{bdryEq} allows us to solve for~$\eta$:
\begin{equation}
  \label{eq:ElimEta}
  \eta = -\frac{1}{4}  (\gamma_5 \nabla_\mu + i  A_\mu) \gamma^\mu\epsilon \ .
\end{equation}

Finally, inserting  this back into \eqref{bdryEq}, we find
\begin{equation}\label{eq:cks0}
	 \nabla^A_\mu  \epsilon_+  = \frac{1}{4}   \gamma_\mu   D^A \epsilon_+ \ ,
\end{equation}
where $\nabla_\mu^A = \nabla_\mu - i  A_\mu$ and $D^A =  \gamma^\mu  \nabla_\mu^A$. 
This is the equation for a charged conformal Killing spinor and will be the starting point of our subsequent analysis.
Note that a similar equation is given for $\epsilon_-$ by complex conjugation.

\subsection{Conformal Killing spinors and superconformal theories} 
\label{sub:csugra}

The equation (\ref{eq:cks0}) was derived using holographic methods but its use is not limited to theories with an holographic dual.
Indeed, the existence of a charged conformal Killing spinor is precisely the condition that allows to preserve supersymmetry for
any superconformal field theory on a curved background.
This has been discussed in detail in \cite{ktz} for the Euclidean case. It works similarly in Lorentzian signature and we will now review the argument.

In order to define a supersymmetric theory on a curved manifold $M$ we can use the strategy of \cite{festuccia-seiberg} which consists in coupling the theory to supergravity and then freeze the fields of the gravitational multiplet. The value of the auxiliary fields determines the coupling of the theory to the curved background. 

The appropriate supergravity for a superconformal theory is conformal supergravity, whose fields are $g_{\mu\nu}$, $\psi_\mu$ and $A_\mu$.
In order to preserve some supersymmetry, the gravitino variation must vanish. With obvious redefinitions it reads \cite{kaku-townsend-vannieuwenhuizen3,vanproeyen-conformal}
\begin{equation}\label{eq:supconf}
\delta \psi_\mu =\left ( \nabla_\mu  - i A_\mu \gamma_5\right )\epsilon  + \gamma_\mu \gamma_5 \eta 
\end{equation}
where $\epsilon$ is the parameter for the  supersymmetries $Q$ and $\eta$ for the superconformal transformations $S$.
We see that the vanishing of (\ref{eq:supconf}) is the same as equation (\ref{bdryEq}) which, in turn,  is equivalent to the CKS equation. 

It is crucial for the argument that 
the algebra of the superconformal transformations of $g_{\mu\nu},\psi_\mu, A_\mu$ closes off shell \cite{das-kaku-townsend}.  Therefore  the 
variation (\ref{eq:supconf})  depends only  on the background field $A_\mu$ and is not modified by the coupling to matter. Moreover, the supergravity action for the fields $g_{\mu\nu},\psi_\mu, A_\mu$ is separately
invariant and can be safely omitted without spoiling the superconformal invariance of the matter part.



In the next section we will discuss what the existence of a conformal Killing spinor 
implies for the geometry of the four-dimensional space-time.

\section{Geometry of conformal Killing spinors in Lorentzian signature} 
\label{sec:cks}

In this section we will analyse the geometrical content of conformal Killing spinors (\ref{eq:cks0}), charged under a gauge field $A$. 
This equation is also known as {\it twistor equation}, and it is well-studied in conformally flat spaces \cite{penrose-rindler-vol2}. The case where $A=0$ has already 
been analysed  in \cite{lewandowski}: all possible spaces on which a conformal Killing spinor exists were classified. 
It turns out that they fall in two classes: Fefferman metrics, and pp-wave spacetimes. We will review these two as particular cases (with $A=0$) 
of our more general classification in section \ref{sub:a0}. 
As stated in the introduction, we will find that a charged conformal Killing spinor exists if and only if there exists a null conformal Killing vector.
To explain the computations that lead to this result, we need to review first some geometrical aspects of four-dimensional spinors in Lorentzian signature.

\subsection{Geometry defined by a spinor} 
\label{sub:geospinor}

In this section we review the geometry associated with a Weyl\footnote{We could also use a Majorana spinor.} spinor $\epsilon_+$  in Lorentzian signature. 
As in section \ref{sec:ads}, we work in the signature $(-,+,+,+)$ and with real gamma matrices. We start with a spinor of positive chirality $\epsilon_+$ and its complex conjugate $\epsilon_-\equiv (\epsilon_+)^*$. We can
use $\epsilon_+$ and $\gamma_\mu \epsilon_-$ to form a basis for the spinor of positive chirality and  $\epsilon_-$ and $\gamma_\mu \epsilon_+$ for those of negative chirality.
A convenient way of choosing the basis is obtained as follows. 
At every point  where  $\epsilon_+$ is not vanishing, it defines a  real null vector $z$ and a complex two form $\omega$.
We can express this fact in terms of  bispinors\footnote{We are using conventions where $\ast   \alpha\wedge \alpha = || \alpha ||^2 {\rm Vol}_4$ and ${\rm Vol}_4 = e^{0123} =  \frac12 e^{+-23}$.}
\begin{equation}\label{eq:bisp}
	\epsilon_+ \otimes\overline {\epsilon}_+ = z + i \ast z
	\ ,\qquad
	\epsilon_+ \otimes \overline{\epsilon}_- \equiv \omega 
	\ ,
\end{equation}
where, as usual,  $\overline{\epsilon} =\epsilon^\dagger \gamma^0$. Equivalently, as spinor bilinears the forms read
\be \label{eq:z}
 z_\mu = \frac{1}{4} \bar\epsilon_+ \gamma_\mu \epsilon_+\, , \qquad 
\omega_{\mu\nu} = -\frac{1}{4}  \bar \epsilon_-  \gamma_{\mu\nu} \epsilon_+ \,.
 \ee
It can be shown easily that
$z\wedge \omega =0$, which implies that we can write 
\begin{equation}\label{eq:zw}
	\omega= z \wedge w
\end{equation}
for some complex one-form $w$. The form $\omega$ looks very similar to the holomorphic top-form of an almost complex structure;
in section \ref{sub:coord} we will make this similarity more precise by introducing the concept of {\it CR-structure}. 
One can then show that the spinor $\epsilon_+$ is annihilated by $z$ and $w$, namely\footnote{When acting on a spinor the dot denotes Clifford multiplication, as in $z \cdot \epsilon = z^\mu \gamma_\mu \epsilon$.}
\begin{equation}\label{eq:ann}
z \cdot \epsilon_+ = w \cdot \epsilon_+ =0\, ,
\end{equation}
and
\begin{equation}
	z^2= z\cdot w = w^2= 0 \ ,\qquad w\cdot\bar w =2\ .
\end{equation}
We can think of $z$ and $w$ as elements of a local frame: $z=e^+$, $w=e_2-i e_3$. In order to complete the frame we 
can introduce another real one-form $e^-$ such that
\begin{equation}\label{eq:e-}
	(e^-)^2=0 \ ,\qquad e^-\cdot z = 2 \ ,\qquad e^- \cdot w =0 \ .
\end{equation}
The four-dimensional metric then takes the form 
\begin{equation}\label{eq:metric}
	ds^2 =  z \,  e^- + w \,  \bar w\ .
\end{equation}

Notice that the pair $(z,\omega)$ is uniquely defined by the spinor, while the frame $\{z,e^-,w,\bar w\}$ is not. This is because, 
given a $w$ that satisfies 
(\ref{eq:zw}), any other one-form of the form $w+\alpha z$ still satisfies it. After having fixed $w$, $e^-$ is uniquely determined by the 
conditions (\ref{eq:e-}). Alternatively, one can pick any null $e^-$ such that $e^-\cdot z=2$; a complex $w$ orthogonal to $e^-$ and $z$ and such that $w^2=0$, $w\cdot \bar w =2$ is then uniquely determined. 

In summary, the vielbein $\{z,e^-,w,\bar w\}$ is not uniquely determined by $\epsilon_+$; rather, it is determined up to the ambiguity
\begin{equation}\label{eq:amb}
	w \to w + \alpha z \ ,\qquad e^- \to e^- - \bar\alpha w - \alpha \bar w - |\alpha|^2 z\ .
\end{equation}
The complex function $\alpha$ has to do with the fact that $\epsilon_+$ by itself describes an $\rr^2$ 
structure\footnote{The stabilizer of the light-like vector $z$ is 
${\rm SO}(2)\ltimes \rr^2$; $w$ breaks the SO(2) to the identity. For more details see \cite{10d}. 
See also section 4 of \cite{gutowski-martelli-reall} for a similar discussion in six dimensions.}, 
rather than the identity structure that would be described by the vielbein $\{z,e^-,w,\bar w\}$.


\subsection{Intrinsic torsions} 
\label{sub:int}

We are now ready to define a basis of spinors. For the positive chirality we can take
\begin{equation}
\label{eq:+basis}
\epsilon_+\ , \qquad e^ - \cdot  \epsilon_-
\end{equation}
and for negative chirality
\begin{equation}
\label{eq:-basis}
\epsilon_-\ , \qquad e^- \cdot  \epsilon_+ \, .
\end{equation} 
It follows from (\ref{eq:ann}) and the previous definitions that
\begin{equation}\label{eq:ge+}
\gamma^\mu \epsilon_+ = - w^\mu \epsilon_- +\frac 1 2 z^\mu e^-\cdot \epsilon_+ \ .
\end{equation}
Using the basis (\ref{eq:+basis}), we can expand
\begin{equation}\label{eq:pq}
	\nabla_\mu \epsilon_+ = p_\mu \epsilon_+ + q_\mu e^- \cdot \epsilon_-\ .
\end{equation}
$p_\mu$, $q_\mu$ are (locally) complex one-forms. 
They can be interpreted as intrinsic torsions for the $\rr^2$ structure defined by $\epsilon_+$.\footnote{For a $G$-structure, one decomposes $\Lambda^2 T = g \oplus k$ (where $g$ is the Lie algebra of $G$); the intrinsic torsion is then given by $k \otimes T$. In our case, $G=\rr^2$, so $k$ is 4-dimensional, and $k\otimes T$ is 16-dimensional. These are precisely the eight complex components of the complex one-forms $p$ and $q$.} 
It is also possible to express $p$ and $q$ in terms of exterior differentials. In order to do so, we can use the auxiliary piece of data $e^-$, which allows to define a vielbein $\{z, e^-, w, \bar w\}$, as described in section \ref{sub:geospinor}. This vielbein is an identity structure. The intrinsic torsion of an identity structure $e^a$ is expressed by the ``anholonomy coefficients'' $c^a{}_{bc}$ defined by $de^a = c^a{}_{bc} e^b \wedge e^c$. As shown in Appendix \ref{app:tor}, we can parametrize the $de^a$ as
\begin{subequations}\label{eq:viel}
	\begin{align}
		d z &=  2 {\rm Re}\, p \wedge z + 4 {\rm Re}(q \wedge \bar  w)\ , \label{eq:dz}\\
		d w &=  - 2 \rho \wedge z + 2 i {\rm Im} p \wedge  w -2 q \wedge e^- \ ,  \label{eq:dw}\\
		d e^- & =  4 {\rm Re}(\rho  \wedge \bar w) -2 {\rm Re} p \wedge e^- \ . \label{eq:de-} 
	\end{align}
\end{subequations}
Here $p$ and $q$ are precisely the one-forms appearing in the covariant derivative of the spinor (\ref{eq:pq}), while $\rho$ is a new one-form which is an intrinsic torsion for the identity structure $\{z, e^-, w, \bar w\}$ but not for the $\rr^2$ structure defined by $\epsilon_+$. In four dimensions, we have $4\times 6=24$ real anholonomy coefficients, 
which we can identify with  three complex one-forms $p,q$ and $\rho$. 

Alternatively, we can extract $p$ and $q$ from the forms $z$ and $\omega$  defined in (\ref{eq:bisp}). 
One can indeed derive the following differential constraints
\begin{subequations}\label{eq:difftorsions}
	\begin{align}
	d z & = 2 {\rm Re}\, p \wedge z + 4 {\rm Re} ( q \wedge \bar w)\, , \label{eq:dzag}\\
	d \omega &= 2 p \wedge \omega - 2 q \wedge ( z \wedge e^- + w \wedge \bar w )\, , \label{eq:domega}\\
	(e^- \cdot \nabla )\, \omega &= 2 (p\cdot e^-) \omega - 2 (q \cdot e^-)  ( z \wedge e^- + w \wedge \bar w ) \, ,
	\end{align}
\end{subequations}
which allow to determine $p$ and $q$ from the geometry. Notice that $dz$ and $d \omega$ alone would not be enough to determine $p$ and $q$.


\subsection{Conformal Killing spinors are equivalent to conformal Killing vectors} 
\label{sub:ckv}

In this section we study the geometrical constraints imposed by the existence of a charged conformal Killing spinor $\epsilon_{+}$.
As discussed in section \ref{sub:csugra}, on the resulting curved backgrounds one can define a superconformal field theory preserving some supersymmetry.
We will show that existence of a charged conformal Killing spinor is equivalent to the existence of a conformal Killing vector. In turn, this allows to introduce local coordinates, in which the metric 
(or equivalently the frame) takes a canonical form, generalizing the one discussed in \cite{lewandowski}, corresponding to $A=0$. 
We shall present this metric in section \ref{sec:nm}, in the case when the conformal Killing vector becomes a Killing vector. The two metrics are  simply related  by a Weyl rescaling. 

Similarly to what is known for uncharged conformal Killing spinors \cite{baum-CKS},
it is straightforward to show that if $\epsilon_+$ is a charged conformal Killing spinor,
the null vector $z_\mu$ defined in $\eqref{eq:z}$ is a conformal Killing vector. We now show that also the opposite is true.

First of all, notice that not only does a spinor $\epsilon_+$ determine a null vector $z$ (via (\ref{eq:bisp}) or (\ref{eq:z})), but also that in a sense the opposite is true. Indeed, let us study the map $\epsilon_+ \mapsto z$. The space of spinors with fixed $\epsilon_+^t \epsilon_+$ is an $S^3$ in the four-dimensional space of all spinors. This is mapped by (\ref{eq:z}) into the space of all null vectors $z$ with fixed $z^0$, which is an $S^2$ (the so-called ``celestial sphere''). This is the Hopf fibration map, whose fibre is an $S^1$. So, to any null vector $z$ one can associate a U(1) worth of possible spinors $\epsilon_+$ whose bilinear is $z$.

Let us now move on to differential constraints. We consider the equation defining a {\it charged} conformal Killing spinor, or twistor-spinor, 
\begin{equation}
\label{eq:cks}
\nabla_{\mu}^{A}\epsilon_{+}=\frac{1}{4}\gamma_{\mu}D^{A}\epsilon_{+} \ ,
\end{equation}
where $\nabla_{{\mu}}^{A}=\nabla_\mu -i A_{\mu}$ and $D^{A}$ is the covariantized Dirac operator $D^{A}=\gamma^{\mu}\nabla_{\mu}^{A}$. 
Note that the equation does not mix chiralities, and we consider the case
of a positive chiral spinor. $A$ is a real connection and  $\epsilon_{+}$ is a section of the U(1) Hopf fibration described in the previous paragraph.

We can expand equation (\ref{eq:cks}) in the basis (\ref{eq:+basis}). Using (\ref{eq:pq}), we obtain  a set of linear equations for  $p$, $q$ and 
the gauge field $A$. Since the gamma-trace of equation (\ref{eq:cks}) is trivial we find a total of six complex constraints 
\begin{equation}
\label{eq:cksconstraints}
\begin{split}
	q \cdot z =0\ ,  & \qquad p^A \cdot e^- =0\ ,  \qquad p^A \cdot z = 2 q \cdot \bar w  \ , \\
	q \cdot w =0 \ ,& \qquad p^A \cdot \bar w =0 \ ,  \qquad p^A \cdot w =  - 2  q \cdot e^- \ ,
\end{split}
\end{equation}     
where $p^A_\mu = p_\mu - i A_\mu$.

Two of these conditions will determine the real gauge field $A$. The remaining eight real conditions are constraints to be imposed on the geometry.
We now show that these constraints are equivalent to the existence of a conformal Killing vector.

A short computation shows that
\begin{equation}\label{eq:nablaz}
\nabla_\mu z_\nu = 2 {\rm  Re} \left ( p_\mu z_\nu + 2 \,\bar q_\mu w_\nu \right )~.
\end{equation}
 
Taking the  anti-symmetric part of this equation we reproduce the first equation in (\ref{eq:viel}). Taking the symmetric part and imposing that $z$ is a conformal Killing vector,
\begin{equation}\label{eq:ckv}
({\cal L}_z g)_{\mu\nu}= 2\nabla_{(\mu} z_{\nu )} = \lambda g_{\mu\nu}\ ,
\end{equation}
we obtain  the conditions
\begin{equation}
	\begin{split}
		{\rm Re}\, p \cdot e^- =0 \ ,& 
		\qquad q \cdot z = q\cdot w =0  \ ,\\
	 {\rm Re}\, p \cdot z = 2 {\rm Re} ( q\cdot \bar w )\ , & \qquad 
	{\rm Re}\, p \cdot w = - q \cdot e^-\ ,
	\end{split}
\end{equation}
 with 
\begin{equation}\label{eq:lambda}
	\lambda \,\equiv\, 4\, {\rm Re} ( q\cdot \bar w )\ .
\end{equation}
This set of eight real conditions is precisely the  subset of the constraints (\ref{eq:cksconstraints}) not involving $A$, as previously stated.

To summarise, we showed that on any manifold $M_4$ with a null conformal Killing vector we can find a charged conformal Killing spinor.
 In section \ref{sub:a} we will  give an expression for the gauge field $A$ under which the conformal Killing spinor is charged.

Notice that the presence of a conformal Killing spinor also implies 
\begin{equation}\label{eq:omegacks}
	d \omega  \,=\,   i\Big( 2 A -  3  * (q \wedge e^- \wedge \bar w)\Big)\wedge \omega  \ .
\end{equation}
In the Euclidean signature case, one finds \cite{ktz} a very similar condition, $d \omega = W\wedge \omega$, where $\omega$ is the $(2,0)$ 
form of a complex structure. While in that case that condition turns out to be necessary and sufficient for the existence of a charged conformal Killing spinor, 
in the present case of Lorentzian signature this condition alone is not sufficient to imply supersymmetry.


\subsection{Conformal Killing spinors are equivalent to conformal Killing--Yano forms} 
\label{sub:ckf}

Our two-form $\omega$ satisfies an interesting property, namely it is a {\it charged} conformal Killing--Yano form (CKF). In general, a $p$-form $\varphi$ on a $d$-dimensional space(-time) $(M,g)$ is conformal Killing--Yano (or\  simply conformal Killing) if it satisfies the equation
\begin{equation}\label{eq:ckfd}
	\nabla_\rho \varphi_{\mu_1 \ldots \mu_p} \,=\, \nabla_{[\rho} \varphi_{\mu_1 \ldots \mu_p]} + \tfrac{p}{d-p+1}\, g_{\rho [\mu_1}\nabla^{\sigma} \varphi_{|\sigma| \mu_2 \ldots \mu_p]}\ .
\end{equation}
This is a conformally invariant equation: if $\varphi$ is a conformal Killing form on $(M,g)$ and the metric is rescaled as $g \to \tilde g = e^{2 f}g $, then the rescaled form $\tilde \varphi = e^{(p+1)f}\varphi$ is conformal Killing on $(M, \tilde g)$.

In the uncharged case ($A=0$), it is known that the bilinears of conformal Killing spinors are conformal Killing forms, see e.g. \cite{semmelmann}. We already saw that this is true for $z$, since a conformal Killing one-form is just the dual of a conformal Killing vector. For a two-form in four dimensions, this is easiest to check in the two-component formalism for spinors. Because of its definition in (\ref{eq:bisp}), $\omega$ can be written as $\omega^{\alpha \beta} = \epsilon^\alpha \epsilon^\beta$, and the CKF equation (\ref{eq:ckfd}) reads
\begin{equation}\label{eq:ckfdot}
	D_{\dot \alpha}^{(\beta} \omega^{\gamma \delta)}= 0 \ .
\end{equation}
The CKS equation reads in this formalism $D_{\dot \alpha}^{(\beta} \epsilon^{\gamma)}=0$, which implies obviously (\ref{eq:ckfdot}). Since for us the CKS is actually charged under $A$, we obtain that $(\nabla^A)_{\dot \alpha}^{(\beta} \omega^{\gamma \delta)}=0$, where $\nabla_\mu^A = \nabla_\mu - 2i A_\mu$; or, going back to four-component language, 
\be \label{eq:ckf}
\nabla^A_\rho \omega_{\mu \nu} \,=\, \nabla^A_{[\rho} \omega_{\mu \nu]} - \tfrac{2}{3}\, g_{\rho [\mu}\nabla^{A\,\sigma} \omega_{\nu] \sigma }\ ,
\ee
which is a charged version of the standard conformal Killing form equation.

It is interesting to ask to what extent this property can be used to characterize our spacetime, similarly to what we saw in section \ref{sub:ckv}. First of all, we should ask when a two-form $\omega$ can be written as a spinor bilinear as in (\ref{eq:bisp}). One possible answer is that the form should define an $\rr^2$ structure; namely, that the stabilizer of $\omega$ in SO(3,1) should be $\rr^2$. We can\  also give an alternative, more concrete characterization by using again the two-component formalism for spinors. The two-form $\omega$ should be imaginary self-dual, which means it is in the $(1,0)$ representation of SO(3,1); the corresponding bispinor then is a symmetric matrix $\omega^{\alpha \beta}$. As a $2\times2$ matrix, this can be factorized as $\epsilon^\alpha \epsilon^\beta$ if and only if it has rank 1, which is equivalent to $\det(\omega)= \frac12 \epsilon_{\alpha \beta} \epsilon_{\gamma \delta} \omega^{\alpha \gamma} \omega^{\beta \delta}=0$; in the original form language, $\omega_{\mu \nu} \omega^{\mu \nu}=0$. So we have obtained that a two-form $\omega$ can be written as a bispinor as in (\ref{eq:bisp}) if and only if it is imaginary self-dual and null:
\begin{equation}
	\omega = \epsilon_+ \otimes \overline{\epsilon_-} \qquad \Longleftrightarrow \qquad 
	\left\{ \begin{array}{cc}
		*\omega= i \omega\ ,\\
		\omega^{\mu \nu} \omega_{\mu \nu} = 0\ . 
	\end{array}\right.
\end{equation} 

Remarkably, it turns out that the content of the equation (\ref{eq:ckf}) for a CKF is exactly the same as the content of the system (\ref{eq:cks}) for a CKS. Indeed, if one uses (\ref{eq:pq}) in (\ref{eq:ckf}) (or in (\ref{eq:ckfdot})), the system one finds is exactly (\ref{eq:cksconstraints}). Thus, we can conclude that a choice of metric and gauge field $A$ admits a charged CKS if and only if it admits a null, imaginary self-dual charged CKF. 

This reformulation is slightly less interesting than the one involving a null CKV in section \ref{sub:ckv}. Although CKF's do have physical applications (such as helping in finding first integrals of the geodesic equation, see for example \cite{walker-penrose}), their geometrical meaning is less compelling than that of a CKV. Moreover, one needs both the data of the geometry and of the gauge field $A$ to check the condition (\ref{eq:ckf}), whereas in the previous section we saw that the presence of a null CKV tells us that a geometry can admit a charged CKS for some $A$ (without having to guess its form, which will actually be determined in section \ref{sub:a}). Last but not least, the CKV condition is computationally easier to check than the CKF condition (\ref{eq:ckf}). 


\subsection{Determining the gauge field} 
\label{sub:a}

The gauge field $A$ can be determined by the four equations in (\ref{eq:cksconstraints}) involving $A$. One possible expression is
\begin{equation}\label{eq:A}
	A= {\rm Im}\big(p + i * (q\wedge e^- \wedge \bar w)\big)\ .
\end{equation}
Here, $p$ and $q$ are intrinsic torsion forms that can be computed for example from (\ref{eq:viel}). 
The fact that (\ref{eq:A}) involves $e^-$ might look puzzling, since, as we stressed in section \ref{sub:geospinor}, $e^-$ is an auxiliary degree of freedom, not one 
determined by the spinor $\epsilon_+$. More precisely, the vielbein $\{z,e^-,w,\bar w\}$ is only defined up to the freedom (\ref{eq:amb}).
 From the definition of $p$ and $q$ in (\ref{eq:pq}), using (\ref{eq:ge+}) we see that under (\ref{eq:amb})
\begin{equation}
	p_\mu \to p_\mu - 2 \bar \alpha q_\mu \ ,\qquad q_\mu \to q_\mu \ .
\end{equation}
Using this and (\ref{eq:A}), we can show that $A$ is invariant under (\ref{eq:amb}). This means it is independent on the choice of $e^-$, and is in fact determined by $\epsilon_+$ alone.

We now show that the gauge curvature is invariant under the action of the vector field $z$, namely that
\begin{equation}
{\cal L}_z F= 0 \, .
\end{equation}
We first compute the Lie derivative of a set of vielbein with respect to the vector $z$. Using equation (\ref{eq:viel}) and the constraints (\ref{eq:cksconstraints}) imposed by the conformal 
Killing spinor equation we find
\begin{align}
\label{eq:Liegen}
{\cal L}_z z &= \iota_z d z= \lambda z \, ,\nonumber \\
{\cal L}_z w & = \iota_z d w = -2 ( {\rm Re} p \cdot w + \rho\cdot z) z + \left( \frac \lambda 2 + i z\cdot A + 3 i {\rm Im}(q\cdot \bar w)\right) w \, ,\\
{\cal L}_z e^{-} & = \iota_z d e^{-} = 2 ( {\rm Re} p \cdot \bar w + \bar \rho\cdot z) w + 2 ( {\rm Re} p \cdot w + \rho\cdot z) \bar w \ , \nonumber
\end{align}
where $\lambda$ has been defined in (\ref{eq:lambda}).
In order to simplify these expressions  we can make use of the freedom in the choice of a basis (\ref{eq:amb}) to set
\begin{equation}
{\rm Re}\, p \cdot \bar w + \bar \rho\cdot z = 0 
\end{equation}
and the gauge invariance to impose
 
\begin{equation} \label{eq:zdotA}
z\cdot A  \,=\, - 3\, {\rm Im}(q\cdot \bar w)\, .
\end{equation}
At this point, the Lie derivative of the vielbein simply 
becomes\footnote{As we will see in section \ref{sec:nm}, in the new minimal case  $\lambda=0$, hence $z$ is a Killing vector and the Lie derivative of the vielbein 
vanishes, ${\cal L}_z z={\cal L}_z w={\cal L}_z e^-=0$. In the notation of section \ref{sec:nm} 
the gauge condition (\ref{eq:zdotA}) reads $a\cdot z=0$.\label{footnote:gauge}}
\begin{equation}
	\label{eq:lie}
	{\cal L}_z z = \lambda z \ ,\qquad
	{\cal L}_z w = \frac \lambda 2 w \ ,\qquad 
	{\cal L}_z e^- = 0 \ ,
\end{equation}
which is consistent with \eqref{eq:ckv}. We can also take the Lie derivative of (\ref{eq:viel}) to compute the Lie derivatives of the torsions
\begin{align}
{\cal L}_z p \,& = \,\frac14 (d\lambda\cdot z) e^{-} + \frac14 (d\lambda\cdot w) \bar w\ , \nonumber\\
{\cal L}_z q \,& =\, \frac \lambda 2 q -  \frac18 (d\lambda\cdot z) w - \frac18 (d\lambda\cdot w) z \, ,\\
{\cal L}_z \rho \,& = \,-\frac \lambda 2 \rho -  \frac18 (d\lambda\cdot w) e^{-} + \frac18 (d\lambda\cdot e^{-}) w\, .
\nonumber
\end{align}
It is then straightforward to check from equation (\ref{eq:A}) that in our gauge ${\cal L}_z A = 0$. It follows that
\begin{equation} \label{eq:zdotF}
\iota_z F \,=\, {\cal L}_z A - d(z\cdot A) \,=\, d(3 {\rm Im}(q\cdot \bar w))\ .
\end{equation}
Notice that this expression is independent of the choice of gauge and frame (due to (\ref{eq:cksconstraints})) and it is valid in general.
It follows from (\ref{eq:zdotF})  that $F$ is invariant, ${\cal L}_{z}F=0$. 



\section{Supersymmetric theories with an R-symmetry} 

\label{sec:nm}

In this section we will discuss an alternative supersymmetry equation, that arises as the rigid limit
of  new minimal supergravity \cite{sohnius-west,sohnius-west2}.   This formulation is particularly well suited to 
describe supersymmetric field theories with an Abelian R-symmetry, and it may be thought 
as a special case of the CKS equation.

\subsection{New minimal supersymmetry equation}


Solutions of the conformal Killing spinor equation (\ref{eq:cks}) are closely related to solutions of the supersymmetry equation 
\begin{equation}\label{eq:nm}
	\nabla_\mu \epsilon_+ = -i \left( \frac12 v^\nu \gamma_{\nu\mu} + (v-a)_\mu\right) \epsilon_+ \ ,
\end{equation}
arising from the rigid limit of new minimal supergravity\footnote{We use lower-case letters $a$ and $v$ for the auxiliary fields of new minimal supergravity, in order to avoid confusion with the $A$ of conformal supergravity we have been using until now.} \cite{sohnius-west,sohnius-west2}.  Here $a$ and $v$ are real vectors and $v$ is required to satisfy $d\ast v=0$. 
When this condition has a solution, we can consistently define supersymmetric field theories on the four-manifold $M_4$, with background fields $v$ and $a$, using 
the strategy of \cite{festuccia-seiberg}.

It is simple to see that a solution of (\ref{eq:nm}) is a conformal Killing spinor associated with the gauge field  $A = a - \frac 32 v$. It follows from our analysis in section \ref{sub:ckv} that there should exist a null conformal Killing vector. It is in fact straightforward to see with a direct computation that equation (\ref{eq:nm}) implies that $z_\mu= \frac 1 4 \overline{\epsilon}_+ \gamma_\mu \epsilon_+$ is not only conformal Killing, but actually even Killing.

Vice versa, if we start  with a solution of  the conformal Killing spinor equation (\ref{eq:cks}) without zeros, charged under a connection $A$, we can define a complex vector $v$ through
\begin{equation}\label{eq:vdef}
	D^A \epsilon_+ \,\equiv\, 2i\, v \cdot \epsilon_+~.
\end{equation}
Every spinor of negative chirality can indeed be written as a linear combination of gamma matrices acting on $\epsilon_+$. If $\epsilon_+$ has no zeros, $v$ is defined everywhere\footnote{Conformal Killing spinors with zeros do exist; see for example \cite{leitner} for a characterization.}. Using (\ref{eq:pq}) we can express some components of $v$ in terms of $q$
\begin{equation}
\label{eq:vp}
w \cdot v \,=\, 2 i\, q \cdot e^- \ , \qquad\qquad z \cdot v \,=\, - 2 i\, q \cdot \bar w \ . 
\end{equation}
All other components of $v$ are immaterial and $v$ itself is not uniquely determined, since we can always add to it a term along $z$ and $w$ (recall that $z \cdot \epsilon_+= w  \cdot \epsilon_+ =0$).
We can use this freedom to make $v$ real, except for an imaginary part given by 
\begin{equation}\label{eq:imv}
	{\rm Im} \,v \,=\, - \frac \lambda 4 e^-\, ,
\end{equation} 
where $\lambda$ was defined in (\ref{eq:lambda}). This rewriting of $q$ in terms of $v$ will be useful in section \ref{sec:bulk}, where we will perform a comparison between bulk and boundary solutions. 
It is now easy to show that (\ref{eq:cks}) can be rewritten as equation (\ref{eq:nm}) with $a = A + \frac 32 v$. 

So far, all we have done is rewriting the equation for conformal Killing spinors (\ref{eq:cks}) as (\ref{eq:nm}); $v$, however, is potentially still complex, with imaginary part given by (\ref{eq:imv}), and in general $d*v \neq 0$. 
We will now show that $v$ can be made real by an appropriate Weyl rescaling $g_{\mu\nu}\to e^{2f} g_{\mu\nu}$. 
To see this, remember that $z$ is a conformal Killing vector, namely a vector satisfying (\ref{eq:ckv}). 
 However, a null conformal Killing vector can always be made a null Killing vector by a Weyl transformation.
In particular,
\begin{equation}
	{\cal L}_z g_{\mu \nu}= \lambda g_{\mu \nu} \ \Rightarrow \ 
	{\cal L}_z (e^{2f}g_{\mu \nu})= (\lambda + 2 z \cdot df) g_{\mu \nu}.
\end{equation}
In coordinates where $z=\frac{\del}{\del y}$, it is then enough to solve $2\frac{\del f}{\del y}= - \lambda$.
This is possible as long as there are no closed time-like curves. In the rescaled metric $e^{2f} g_{\mu\nu}$, $z$ is now a Killing vector, which implies that $\lambda=0$; from (\ref{eq:imv}), we then see that $v$ is real.
Moreover, a similar argument shows that we can use the remaining ambiguity in shifting the $z$ component of $v$ to arrange for 
\begin{equation}\label{eq:d*v}
	d * v = 0 \ .
\end{equation} 

Hence we have shown that, by a conformal rescaling of the metric, one can take the charged conformal Killing spinor equation (\ref{eq:cks}) to the 
condition of unbroken supersymmetry in new minimal supergravity (\ref{eq:nm}). 
The fact that one can bring (\ref{eq:cks}) to (\ref{eq:nm}) was to be expected because of the formalism of conformal 
compensators (for a review see \cite{vanproeyen-conformal}). In that formalism, one obtains new minimal supergravity by coupling a tensor multiplet to conformal supergravity, and by then giving an expectation value to the tensor multiplet. 

To summarize, the geometrical constraints imposed by the new minimal equation just amounts to the existence of a null Killing vector
 $z$. As a check, we can count components. 
The new minimal equation (\ref{eq:nm}) brings 16 real constraints and  the existence of a Killing vector brings 9 real conditions. 
The remaining 7 real constraints can be used to determine the components of the gauge fields: 4 for $a$ and $3$ for $v$. 
$a$ and $v$ can now be computed as follows. $v$ can be computed from (\ref{eq:vp}), and from (\ref{eq:d*v}), while $a= A + \frac 32 v$, where $A$ was given in (\ref{eq:A}). 
Recall that the intrinsic torsion $p$ and $q$ can be computed for example from (\ref{eq:viel}).

 Finally, we observe that a solution $\epsilon_+$ of the new minimal equations is defined up to a multiplication by a complex number. We can form two independent Majorana spinors
$\epsilon_1= \epsilon_+ + \epsilon_-$ and $\epsilon_2= i (\epsilon_+ -   \epsilon_-)$ corresponding to two independent real supercharges.  
The commutator of these two supersymmetries closes on the isometry generated by $z$:
\begin{equation}
\label{eq:superalgebra}
[ \delta_{\epsilon_1}, \delta_{\epsilon_2}]   \Phi = {\cal L}^{a}_{z} \Phi\ ,
\end{equation}
where $\Phi$ is any field in the Lagrangian and the Lie derivative ${\cal L}^a$ is covariantized with respect to $a$. The superalgebra can be easily extracted from 
the transformation rules of matter fields in the new minimal supergravity or from the algebra of local  supergravity transformations \cite{sohnius-west, sohnius-west2}.

\subsection{Introducing coordinates} 
\label{sub:coord}

We can obtain more explicit expressions for $v$ and $A$ after introducing a local set of coordinates, as promised earlier. 
First, notice that, using (\ref{eq:vp}), (\ref{eq:imv}) and the fact that $\lambda=0$, equations (\ref{eq:dz}) and (\ref{eq:omegacks}) simplify considerably:
\begin{subequations}\label{eq:dbispnm}
	\begin{align}
		\label{eq:dznm}
		d z &= -2 \iota_v *z\ , \\
		\label{eq:donm}
		d\omega & = 2 i a \wedge  \omega \ . 
	\end{align}
\end{subequations}
As noticed at the end of section \ref{sub:ckv}, the second of these equations is similar to the equation that in Euclidean 
signature implies that the manifold is complex.
(\ref{eq:dbispnm}) can be  used to compute all the components of $a$ and $v$ not along $z$. 
In particular,  (\ref{eq:dznm}) can be inverted to give
\begin{equation}
v^\perp \, \equiv v\, - \tfrac 12 (e^- \cdot v) z \,=\, -\frac 14 \iota_{e^-} *dz\,.
\label{vtrans}
\end{equation}
 As discussed before,  the component of $v$ along $z$ is ambiguous and is determined by requiring (\ref{eq:d*v}).

Given a null Killing vector, there exists a set of natural coordinates adapted to this. 
We will follow the discussion in \cite{gutowski-martelli-reall}. We can introduce a coordinate $y$ such that as vector field 
\bea
z & = & \frac{\de}{\de y}~,
\eea
and then the vector field dual to the one-form $e^-$ introduced earlier can be parameterized as
\bea
e^- &=& 2H \left( \frac{\partial\,}{\partial u} - \mathcal F \frac{\partial\,}{\partial y} \right),
\eea
for some $H$ and $\mathcal F$. Taking as coordinates on the four-dimensional space $(y,u,x^m)$, the functions 
$H$ and $\mathcal F$ do not depend on $y$, and are otherwise arbitrary functions of $u$ and $x^m$. 
In these coordinates, the four-dimensional metric can be written as
\be
ds^2 \,=\, 2H^{-1} (du + \beta)\Big(dy + \noname + \mathcal F (du + \beta) \Big) + H h_{mn} dx^m dx^n,
\label{localmetric}
\ee
where $h_{mn}$ is a two-dimensional metric, while $\beta = \beta_m dx^m$ and $\noname = \noname_m dx^m$ are one-forms. 
Everything depends on $u$ and $x^m$, but not on $y$.
Therefore,  as one-forms,
\bea
z \,=\, H^{-1} (d u + \beta)\,, \qquad e^- \,=\, 2\left( dy + \noname + \mathcal F H z\right) \,.
\label{expframe}
\eea
The remaining elements of the vielbein can be complexified as $w = e^2 - i e^3$. 

Our four-dimensional manifold $M_4$ can be seen as an $\rr$ fibration (with coordinate $y$)
over a three-dimensional manifold $M_3$ (spanned by $\{z,w\}$).
 The latter admits a {\it CR structure}: namely, a one-dimensional complex subbundle 
$T_{1,0}\subset T M_3$, such that $T_{1,0}\cap \overline{T_{1,0}}=\{ 0\}$. Roughly speaking, 
this can be thought of as a complex structure on two of the three dimensions of $M_3$. For 
us, $T_{1,0}$ is spanned by the vector dual to the one-form $\bar w$. From a dual point of view,
the subbundle of $T^* M_3$ spanned by one-forms which are orthogonal to $\overline{T_{1,0}}$ 
has dimension two, and it is spanned by the one-forms 
we have been calling $z$ and $w$; so its volume form is $z\wedge w$, which is the form we have been calling $\omega$, and which in a sense
can be used to characterize the CR structure. The role of $z \wedge w$ actually becomes clearer in higher odd dimension $2n+1$; the 
bundle $T_{1,0}$ is now an $n$-dimensional bundle which should be closed under Lie bracket (just like for a complex manifold). 
The subbundle of $T^* M_3$ orthogonal to $\overline{T_{1,0}}$ now has dimension $n+1$, and it is spanned by forms $z$, $w_1,\ldots,w_n$. 
Integrability of $T_{1,0}$ is equivalent to the statement that 
$d(z\wedge w_1\wedge \ldots \wedge w_n)= a \wedge (z\wedge w_1 \wedge \ldots \wedge w_n)$ 
for some one-form $a$.  Summing up, on our three-dimensional manifold $M_3$ 
the form $z \wedge w$ is the analogue of a holomorphic volume form for a CR-structure, and can be used
to characterize it.

Let us now present expressions for $v$ and $A$ in these coordinates. Evaluating (\ref{vtrans}) we find
\be\label{formulavperp}
v^\perp  = \frac 14 H^{-2} \left[*_2( \beta \wedge \partial_u \beta -d_2 \beta)\right]  e^-  +\frac 12 H *_2 
\left[ \partial_u (H^{-1}\beta) -d_2( H^{-1}) \right] \, ,
\ee
where we defined $d_2 = dx^m \partial_m$ and $*_2$ is the Hodge star operator with respect to the metric $h_{mn}$. 
Inserting   $a\equiv a^\perp + \tfrac{1}{2}(a\cdot e^-) z$  into (\ref{eq:donm}), we determine $a^\perp$ as 
\be
\label{aperp}
a^\perp \, =\,  \frac{1}{4} *_2 \left[d_2 (H^{-1} \bar w)  - \de_u (H^{-1} \beta \wedge \bar w)\right]  w + \mathrm{c.c.}\, ,
\ee
where c.c.~denotes the complex conjugate\footnote{Comparing with  section \ref{sub:a}  we see that $a\cdot z=0$ is a consequence of the Lie 
derivative ${\cal L}_z$ of our vielbein being  zero, and  corresponds to  the gauge condition (\ref{eq:zdotA})  (see footnote \ref{footnote:gauge}).}.
The remaining component of the gauge field is given by $ a\cdot e^- = A\cdot e^- + \frac32 v\cdot e^-$. As already noticed the component $v\cdot e^-$ is ambiguous;   
$A\cdot e^-$ can  be extracted for example from the second and third equations in (\ref{eq:vielnm}) and reads 
\be
\label{genaz}
A \cdot e^- \, = \, \frac{1}{2}H^{-1}*_2 \left[ d_2 \noname + \mathcal F d_2 \beta  + 
(\de_u \noname +  {\cal F}\de_u \beta)\wedge \beta + H \, \mathrm{Re}( \bar w \wedge \de_u w) \right]~.
\ee

Perhaps it is worth emphasizing that the data entering in the metric (\ref{localmetric}), gauge field $A$ and $v$ 
(${H}, {\cal F}, \beta, \noname, h_{mn}$) are completely arbitrary. This is in stark contrast with the typical situation 
in supergravity, where e.g. the Bianchi identities and equations of motion impose more stringent constraints on the geometry.


\subsection{Non-twisting geometries} 
\label{sub:nontw}

In the special case that $z\wedge dz=0$ everywhere, $z$ is hypersurface orthogonal, in the sense that the distribution defined by vectors orthogonal to $z$ is integrable (by Frobenius theorem).  As we will show in section \ref{sec:ex}, 
this corresponds to the case where the Killing vector in the bulk is \emph{null}. 
Since $z$ is hypersurface orthogonal, there exist preferred functions $H$ and $u$ such that
\be                                        
z = H^{-1} du\ .                           
 \ee   
Comparing with equation (\ref{expframe}), we see that in these particular coordinates $\beta=0$. 
After performing a further local change of coordinates to eliminate $\noname$, the metric can be brought to 
the pp-wave form, namely
\be
ds^2 \,=\, 2H^{-1}du \left(dy + \mathcal F du  \right) + H h_{mn} dx^m dx^n~.
\label{pipimetric}
\ee
In addition we have
\bea
\label{eq:vnt} v^\perp  & = &  - \frac 12 H *_2  d_2( H^{-1}) \\
\label{eq:apnt} a^\perp & = & \frac{1}{4}\left[ *_2  d_2 (H^{-1} \bar w) \right]\, w + \mathrm{c.c.}\\
\label{eq:Aznt} A \cdot e^-  &=& \frac{1}{2}*_2 \left[ \mathrm{Re}( \bar w \wedge \de_u w) \right]
\eea
where in particular notice $v \cdot z=0$.


\subsection{The case $A=0$} 
\label{sub:a0}

It is interesting to study what happens in the particular case $A=0$. Actually, as we showed in this section, every solution to the new minimal 
equation (\ref{eq:nm}) is also a solution to the CKS equation (\ref{eq:cks}), and hence must be included in the classification of uncharged conformal 
Killing spinors obtained in \cite{lewandowski}. We will first consider the case $z\wedge dz \neq 0$, and then the case $z\wedge dz=0$.

When $z\wedge dz\neq0$, $z$ is a {\it contact form} on the three-dimensional manifold $M_3$, spanned by $\{z,w\}$.
It follows from (\ref{eq:dznm}) that $\frac{1}{2}z\cdot v \equiv v_- \neq 0$; using (\ref{eq:amb}), 
we can then make $w\cdot v=0$, so that\footnote{As pointed out after (\ref{eq:vp}), the component $v_z$ is immaterial, and can be used to set $d *v=0$.}
\begin{equation}\label{eq:v-z}
	v = v_- e^- + v_z z \ . 
\end{equation} 

Since $A=0$, we have that $a= A + \frac 32 v= \frac32 (v_- e^- + v_z z)$. Moreover, from (\ref{eq:zdotF}) and (\ref{eq:vp}), we see that $v_-$ is actually constant. Our (\ref{eq:dbispnm}) now become
\begin{subequations}\label{eq:fefferman}
	\begin{align}
		\label{eq:dzCR}
		dz & = 2 i v_- w \wedge \bar w \ ,\\
		\label{eq:domCR}
		d \omega & = 3 i v_- e^- \wedge \omega \ .
	\end{align}
\end{subequations}
Moreover, from (\ref{eq:vielnm}) and the fact that ${\rm Re} ( \sigma\cdot \bar w) =0$ as a consequence of (\ref{eq:dzCR}),  we also have
\begin{equation}\label{eq:feffermanbis}
e^- \wedge dw \wedge \bar w = \frac18 (\iota_w \iota_{\bar w} de^-) e^-\wedge z\wedge w \wedge \bar w \, . 
\end{equation}
A metric of the form (\ref{eq:metric}), such that (\ref{eq:fefferman}) and  (\ref{eq:feffermanbis}) hold, 
is called a \emph{Fefferman metric} \cite{fefferman}. It has the property that, 
if one rescales the one-form $z \to \tilde z = e^{2 \lambda} z$, where $\lambda$ is a
 function on the CR manifold $M_3$, and one computes new $\tilde w$, $\tilde e^-$ so that (\ref{eq:fefferman}) and (\ref{eq:feffermanbis}) are still satisfied, 
the new metric $\tilde z \tilde e^- + \tilde w \overline{\tilde w}$ is equal to $ e^{2 \lambda}(z e^- + w \bar w)$.\footnote{Such a characterization of Fefferman metric can be found in  \cite{lee-fefferman}. The metric is defined there by \cite[(3.7)]{lee-fefferman}. The term $L_\theta$ in that equation is 
defined composing (\ref{eq:dzCR}) with the complex structure associated to $T_{1,0}$, the one-dimensional complex subbundle of $T M_3$ defining 
the CR structure on $M_3$, and corresponds to our term $w\bar w$ in (\ref{eq:metric}). Moreover, the term $2 \theta \sigma$ in \cite[(3.7)]{lee-fefferman} 
is identified with the term $e^- z\ $ in our (\ref{eq:metric}), once we compare our (\ref{eq:domCR}) with \cite[(3.5)]{lee-fefferman}.
Finally, \cite[(3.6)]{lee-fefferman} is (\ref{eq:feffermanbis}).} Notice that (\ref{eq:fefferman}) are very similar to the conditions for 
Sasaki--Einstein manifolds (which have Euclidean signature rather than Lorentzian, and odd dimension rather than even). The fact that we found a 
Fefferman metric in the $A=0$, $z\wedge dz \neq 0$ case is in agreement with the classification in \cite{lewandowski}.

Let us  now consider the case $z\wedge dz=0$. Using (\ref{eq:vnt}) and (\ref{eq:apnt}) it follows that $A^\perp=0$ implies  
\begin{equation}
	d_2(\sqrt{H} w) = 0 \ .
\end{equation}
Hence, we can choose a complex coordinate $\zeta$ and a function $\alpha$ so that locally $\sqrt{H} w = d \zeta + \alpha du $. We can then rearrange (\ref{pipimetric}) as $ds^2= H^{-1} [ du ( 2 d y + d \zeta \bar \alpha + d\bar \zeta \alpha + {\cal F} du) + d \zeta d \bar \zeta]$, after suitably redefining ${\cal F}$. Moreover, from (\ref{eq:Aznt}) we learn that the component of $d\alpha$ along $w$ is real. This implies in turn that the one-form $d \zeta \bar \alpha + d\bar \zeta \alpha$ is closed, up to terms $du\wedge (\ldots)$; locally, we can then write $d \zeta \bar \alpha + d\bar \zeta \alpha = d f + g d u$, for some functions $f$ and $g$. We can now further redefine $y$ and ${\cal F}$ to obtain  
\begin{equation}
	ds^2 = H^{-1}[du \left( 2 dy + \mathcal F du  \right) + d\zeta d\bar \zeta] \ ,
\end{equation}
which agrees (locally) with the classification in \cite[Eq.~(41)]{lewandowski}.



\section{Boundary geometry from the bulk} 
\label{sec:bulk}

The general analysis of the supersymmetry conditions in the minimal gauged supergravity in five dimensions  was performed in \cite{gauntlett-gutowski}. 
Here we would like to asymptotically expand these results, to extract a set of conditions on a four-dimensional boundary geometry.
Not surprisingly, at leading order we find agreement with the conditions that we derived from the CKS equation on the boundary in sections \ref{sec:cks} and \ref{sec:nm}.

\subsection{Asymptotic expansion of the bilinears}

The analysis in \cite{gauntlett-gutowski} uses the following set of five-dimensional bilinears:
\begin{equation}
	\begin{split}
		f \epsilon^{IJ} \,&=\, i  \bar \epsilon^I \epsilon^J\ ,\\ 
		V_\alpha \epsilon^{IJ} \,&=\, \bar \epsilon^I \gamma_\alpha \epsilon^J \ , \\ 
		X^{(1)}_{\alpha\beta} + i X^{(2)}_{\alpha\beta} \,&=\, -i  \bar\epsilon^1 \gamma_{\alpha\beta} \epsilon^1 \,=\, -(i  \bar\epsilon^2 \gamma_{\alpha\beta} \epsilon^2)^*  \ , \\ 
		X^{(3)}_{\alpha\beta}\, &= \, \bar\epsilon^1 \gamma_{\alpha\beta} \epsilon^2 \,=\, \bar\epsilon^2 \gamma_{\alpha\beta} \epsilon^1\,.
	\end{split}
\label{allbilinears}
\end{equation}
Here, $f$ is a real scalar, $V$ is a real one-form, and $X^{(i)}$, $i=1,2,3$, are real two-forms\footnote{They satisfy a set of  algebraic relations that can be found in equations (2.8)--(2.12) of \cite{gauntlett-gutowski} (changing the sign of the metric in order to take into account the opposite choice of signature).}. We will also define  $\Omega =  X^{(2)} + i X^{(3)}$.

We can  expand  the bulk bilinears  \eqref{allbilinears} near the boundary using (\ref{eq:expansion}).
In order to facilitate the comparison with the boundary results,
let us again define a complex one-form $v$ via the covariant derivative of $\epsilon$ as in $\eqref{eq:vdef}$, which when plugged into $\eqref{eq:ElimEta}$ yields
\be\label{defv}
\eta = i\gamma^\mu{\rm Re }(v_\mu \epsilon_+) = \frac{1}{2}\gamma^\mu \left( i{\rm Re} v_\mu \epsilon - {\rm Im} v_\mu  \gamma_5 \epsilon \right),
\ee
where we have used that the Majorana condition on $\epsilon$ implies that $\epsilon_+^{\ast} = \epsilon_-$.
Recall that $v$ has a complex part given by (\ref{eq:imv}).
 Recall also the definition of  the boundary bilinears \eqref{eq:z},
which together with the Hodge dual of $z$ correspond to the four-dimensional bilinears 
defined by a single chiral spinor $\epsilon_+$, and determine an $\mathbb{R}^2$
structure.  Note also that using the properties of the 4d gamma matrices, one can check that
\bea\label{dualityomega}
* \omega \,=\, i\, \omega 
\eea
where the Hodge star is four-dimensional and the metric is the boundary metric $ g_{\mu\nu}$.

With these definitions, it is straightforward to compute the asymptotic expansion of the bulk bilinears $\eqref{allbilinears}$
at leading order in $r$, namely
\begin{subequations}  \label{eq:BulkBilAsymp}
\bea
f /8 & \sim &  - \re v \cdot z\ , \label{eq:fasympt}\\
\ell^{-1} V/8 & \sim &  r^2 z +   r^{-1} \im v \cdot z \ d r\ ,
\label{vplug} \\
\ell^{-2} X^{(1)} /8 & \sim &  r dr \wedge z  - r^2 \left({\rm Re}\, \iota_v  * z + \im v \wedge z \right) \ , \label{x1plug}\\
 \ell^{-2} \, \Omega /8  & \sim & i r^3 \omega  + dr \wedge   \iota_v \omega   \ .
\eea
\end{subequations}
Using \eqref{dualityomega} and the identity $ * \iota_v  \omega = i v \wedge \omega$, 
one also finds that at leading order
\bea 
\ell^{-3} \, \hat * \, \Omega/8  & \sim & -i r^3 v \wedge\omega -  r^2 dr \wedge\omega \ ,
    \label{starbarOmega}
\eea
where $\hat *$ denotes the five-dimensional Hodge star.

\subsection{Differential conditions from the bulk}
\label{sub:diffbulk}

The conditions for the existence of  supersymmetric solutions in the bulk  can be written in terms of a set  of differential conditions on the bilinears \cite{gauntlett-gutowski}. We will now expand these conditions near the boundary where the metric is given by (\ref{feffer}) and the gauge field by $\eqref{AnsatzA}$. They read
\begin{subequations}  \label{eq:SUSYbulk}
  \bea 
 df & = & -\frac{2}{3} i_V {}F \label{eq:df}   \\ [2mm]
\hat \nabla_\alpha V_\beta &= & \ell^{-1} X^{(1)}_{\alpha \beta}  + \cdot\cdot\cdot
\label{eq:dV}  \\ [2mm]
\hat \nabla_\alpha X^{(1)}_{\beta\gamma} & = &  2\ell^{-1} \, \eta_{\alpha [\beta} V_{\gamma]}   \label{nablaX1} + \cdot\cdot\cdot\\ [2mm]
\hat \nabla_\alpha \Omega_{\beta\gamma} & = & -i\ell^{-1} \left( 2 \sqrt 3\, \hat A_\alpha \Omega_{\beta\gamma}  + (* \Omega)_{\alpha\beta\gamma} \right)   + \cdot\cdot\cdot \label{nablaOmega} 
\eea
\end{subequations}
where we omitted  terms containing $F$, whenever they are manifestly sub-leading in $r$.\footnote{Full expressions can be found in eqs. (2.15), (2.17), (2.18), (2.19) in \cite{gauntlett-gutowski}.}

We can now further  expand the  bulk differential conditions $\eqref{eq:SUSYbulk}$ near the boundary.
In the computation, we will need the expressions of the Christoffel symbols for the five-dimensional metric with expansion given in \eqref{feffer}. 
We have the following identities
\bea
 \hat \Gamma_{rr}^{\;\mu} = \hat \Gamma_{\mu r}^{\; r} = 0 ~, \qquad \quad \hat \Gamma_{rr}^{\; r} = - \frac{1}{r} ~,
\eea
as well as the expansions
\bea
\hat \Gamma_{\mu\nu}^{\; \rho} \, = \,  \Gamma_{\mu\nu}^{\; \rho} +  \mathcal{O} (r^{-1}) ~, \qquad
\hat \Gamma_{\mu r}^{\;\nu} \, = \,  \frac{1}{r}\delta_\mu^\nu + \mathcal{O} (r^{-1}) ~, \qquad
\hat \Gamma_{\mu\nu}^{\; r}  \,=\,  - r^3  g_{\mu\nu} + \mathcal{O} (r^2) ~,
\eea
where $ \Gamma_{\mu\nu}^{\; \rho}$ denotes the Christoffel symbols of the four-dimensional metric $g_{\mu\nu}$. 
Let us start with equation $\eqref{eq:dV}$. Its symmetric part is simply
\bea
\hat \nabla_{(\alpha} V_{\beta)} & = & 0~,
\label{bulkill}
\eea
which states that $V^\alpha$ is a Killing vector in the bulk. 
It is an easy check to see that, at leading order in $r$, this just says that $z^\mu$ is a 
conformal Killing vector on the boundary.
To this end,  note that the equations having components along $r$ do not give rise to any conditions on the boundary, 
while the ones without leg along $r$ imply
\be
\nabla_{(\mu} z_{\nu)}  \,=\, - g_{\mu \nu} \im v \cdot z  \,\equiv \,\frac{\lambda}{2} \, g_{\mu \nu} \,.
\ee
This is the same condition that we found from the purely four-dimensional analysis in $\eqref{eq:ckv}$ and $\eqref{eq:imv}$.

Having reproduced the existence of a boundary conformal Killing vector from the gravity analysis,
let us now consider the other differential conditions. 
We have reformulated the conditions on the boundary geometry 
in $\eqref{eq:cov}$ and $\eqref{eq:covomega}$ in such a way to make the comparison with the bulk analysis of this section most 
straightforward. 
Plugging \eqref{vplug} and \eqref{x1plug} into the anti-symmetric part of $\eqref{eq:dV}$,
we find that again the only non trivial information at leading order in $r$ comes from the four-dimensional part.
We get the condition
\begin{equation}
\label{pippo}
	d z  =  -2 ({\rm Re}\, \iota_v  *z +{\rm Im} \, v \wedge z)\ ,
\end{equation}
which is just the anti-symmetric part of $\eqref{eq:cov}$. 
Next on the list are equations $\eqref{nablaX1}$ and $\eqref{nablaOmega}$.
At leading order, 
\eqref{nablaX1} and the $(\mu \nu 5)$-part of $\eqref{nablaOmega}$ do not give any new information. 
On the other hand, upon using \eqref{starbarOmega}, the four-dimensional part of equation \eqref{nablaOmega} yields 
\bea
\left( \nabla_\rho - 2 i   A_\rho \right) \omega_{\mu \nu} & = &   i(v \wedge \omega)_{\mu\nu\rho} + i \left( g_{\rho \nu} \omega_{\mu \sigma} -   g_{\rho \mu} \omega_{\nu\sigma}\right) v^\sigma \,,
\eea
which is precisely the equation $\eqref{eq:covomega}$.

The final  equation $\eqref{eq:df}$ would seem to be more problematic. It involves the scalar bilinear in the bulk 
which has no correspondence in the boundary and involves the curvature of the gauge field which we usually neglected because sub-leading in $r$.  However, equation $\eqref{eq:df}$ expands to
\bea
df = -\frac{16}{3} \iota_z  F \, ,
\eea
a relation that we also found on the boundary. It corresponds indeed to  $\eqref{eq:zdotF}$, as we can see by  using  $\eqref{eq:vp}$ and $\eqref{eq:fasympt}$.

We have thus shown that, as expected,  all the conditions for supersymmetry in the bulk  reduce to  conditions that can be derived from the CKS equation on the boundary. In other words, any supersymmetric bulk solution that can be written asymptotically in the Fefferman--Graham form (\ref{feffer}) and with a gauge field $A$ satisfying $\eqref{AnsatzA}$ reduces to the boundary to a metric with a null conformal Killing vector. This vector is  associated with a conformal Killing spinor $\epsilon_+$ charged under $A$. Vice versa,  any Lorentzian metric with a null conformal Killing vector gives rise to a bulk metric (\ref{feffer}) that solves, at leading order, the supersymmetry
conditions of gravity. In this regard, we expect to be able to find a supersymmetric bulk solution with a given  boundary condition order by order in $r$, in the spirit of the  Fefferman--Graham construction. It is then a very hard problem to determine
which boundary metrics give rise to regular solutions in the bulk. Few examples are known in the literature
and they will be reviewed in the next section.

\section{Time-like and null solutions in the bulk}
\label{sec:ex}

In this section we will analyse in more detail
 the classification of supersymmetric solutions of minimal five-dimensional gauged supergravity given in \cite{gauntlett-gutowski}. We will demonstrate how to extract the
boundary data from a bulk solution and we will discuss how the examples found in  \cite{gauntlett-gutowski}
fit in the general discussion of supersymmetric boundary geometries.

Let us analyse some general features of the bulk solutions that can be written asymptotically in the Fefferman--Graham 
form (\ref{feffer}) and with a gauge field $A$ satisfying $\eqref{AnsatzA}$.  
As discussed in the previous section, the five-dimensional vector $V$ is Killing and its asymptotic expansion (written here as the dual one-form), 
\begin{equation}\label{eq:vexpr2}
V  \sim   r^2 z +   r^{-1} \im v \cdot z \ d r + \cdots \, , \qquad\qquad \im v \cdot z =-\frac12 \lambda \ ,
\end{equation}  
gives rise to a null conformal Killing vector $z$ on the boundary.
As in \cite{gauntlett-gutowski}, we can introduce a coordinate $y$ such that
\begin{equation}
V = \frac{\partial}{\partial y} \, .
\end{equation}
In this  (particularly natural) coordinate system  the metric is independent of $y$ and so will be the boundary metric.
This means that $z$ is actually Killing and we can identify the bulk coordinate $y$ here with the coordinate $y$ introduced in section \ref{sub:coord}.
We also learn that the term $ \im v \cdot z =-\frac12 \lambda$, which controls the failure of $z$ at
 being Killing (recall (\ref{eq:imv})), must vanish and
\begin{equation}\label{eq:vexpr22}
V  \sim   r^2 z + \cdots \, .
\end{equation}  
There is no loss of generality here. As discussed  in section \ref{sec:nm}, one can always make $z$ Killing by a Weyl rescaling of the boundary metric. But Weyl rescalings in the boundary are part of diffeomorphisms in the bulk and they can be arranged with a suitable choice of coordinates. In a different coordinate system, for example with an (unnatural) choice of radial coordinate depending on $y$, we would find that $z$ restricts to a conformal Killing vector on the boundary.

The boundary data can be easily extracted from the bulk metric. It follows from our discussion that
the natural framework where to discuss the boundary supersymmetry is that of the new minimal equation. 
 The boundary metric and gauge field $A$ can be read off from equations (\ref{feffer}) and $\eqref{AnsatzA}$. To have full information about the supersymmetry
 realised on the boundary we also need the vector $v$. 
This is real and satisfies equation (\ref{eq:dznm}). It can easily be computed strarting from $z$, using for example (\ref{vtrans}); the component of $v$ along $z$ is ambiguous and is determined by requiring (\ref{eq:d*v}). We will see explicit examples of this procedure in the following.

Note that while $z$ is always null with respect to the boundary metric, 
the five-dimensional Killing vector $V$ can be null or time-like \cite{gauntlett-gutowski}. This follows from the algebraic constraint (equation (2.8) in \cite{gauntlett-gutowski})
\begin{equation}
V^2 = -f^2\ .
\end{equation}
For  $f\ne 0$, $V$ is time-like while, for $f=0$, $V$ is null. The time-like and null solutions have different properties
and, following \cite{gauntlett-gutowski}, we will discuss them separately. Recall from (\ref{eq:fasympt}) that 
\begin{equation}
f \sim - 8 v\cdot z \ ,
\end{equation}
so the time-like and null bulk solutions correspond to $v\cdot z \ne 0$ and $v\cdot z=0$, respectively.
As already noticed in section \ref{sub:nontw}, these correspond to $ z \wedge dz \ne 0$ and  $ z \wedge dz = 0$, respectively. 
It then follows that the null bulk case corresponds to the  non-twisting geometries discussed
in section \ref{sub:nontw}. In the following, we consider these cases in turn, also discussing two explicit examples as an illustration of our general results.

\subsection{Time-like case} 
\label{sub:time}

In the time-like case, the bulk metric can be written as a time-like fibration over a four-dimensional base $B_4$, as 
\begin{equation}\label{eq:bulktimelike}
ds^2 = - f^2 (d y + \tau)^2 +f^{-1} ds^2(B_4)\ ,
\end{equation}
where $f$, $\tau$ and $ds^2(B_4)$ do not depend on $y$.  As a one-form,  $V$ reads
\begin{equation}
V = -f^2 (d y +\tau )\ .
\end{equation}
Supersymmetry in the bulk requires  
the base  $B_4$ to be K\"ahler \cite{gauntlett-gutowski}; in particular, as shown in \cite{gauntlett-gutowski}, this is equivalent to the equations\footnote{These equation are valid in the gauge
$\iota_V \hat A = -\frac{\sqrt{3}}{2} f$,  used in \cite{gauntlett-gutowski}. (\ref{lastminute}) are  therefore  equations on the base $B_4$.} 
\bea
d X^{(1)} \ = \ 0 \, ,  \qquad  d\Omega \ = \ -i\ell^{-1} (2 \sqrt{3} \hat A - 3f^{-1} V)\wedge \Omega\, .
\label{lastminute}
\eea

We are interested in metrics that can be written in the  Fefferman--Graham form (\ref{feffer}). As already discussed, 
such metrics have the properties that at large $r$, $f$ is independent of $r$ and $V\sim r^2 z$. It then follows that $\tau = O(r^2)$.   
The boundary supersymmetry is determined by the background fields $a$ and $v$. $v$ is extracted from  (\ref{eq:dznm}), while $a = A +\frac32 v$,
where $A$ can be read off from $\eqref{AnsatzA}$. Once again, we can check that $v$ is real in such solutions. Indeed, 
a non-vanishing term $r^{-1} \im v \cdot z \,dr$   in (\ref{eq:vexpr2})  
would contradict the mutual consistency of the two metrics  (\ref{eq:bulktimelike}) and (\ref{feffer}) by introducing $dydr$ terms.

It is interesting to write explicitly the asymptotic K\"ahler structure $(X^{(1)},\Omega)$ on the base manifold $B_4$.
Using the freedom to take $v=v_-e^- + v_z z$, combining (\ref{eq:BulkBilAsymp}) with (\ref{lastminute}) we get
\bea
\ell^{-2} X^{(1)}  &  \sim &   r dr \wedge z  - 2 v_- r^2 w \wedge \bar w  = \tfrac{1}{2} d\left(r^2 z\right) \, ,\label{kcones}\\
\ell^{-2} \Omega & \sim &  (2 v_- dr + i r^3 z)\wedge w\, ,
\eea
and correspondingly the K\"ahler metric reads
\bea
\ell^{-2} ds^2 (B_4) & \sim & 2v_-\left( \frac{dr^2}{r^2} + r^2 w \bar w\right) + \frac{1}{2v_-}r^4z^2\, .
\label{asinometric}
\eea
Eq.~(\ref{kcones}) characterises K\"ahler cones, however the asymptotic metric is not homogeneous in $r$, and this is 
reflected by the  $(2,0)$-form $\Omega$. Equations (\ref{kcones})--(\ref{asinometric}) may be thought of as boundary conditions that a K\"ahler base $B_4$ should satisfy.  We also note that on surfaces of constant $r$, $\Omega$ pulls back to a form proportional 
to $z\wedge w$, which characterizes the CR structure on $M_3$, as we saw in section \ref{sub:coord}.


In \cite{gauntlett-gutowski,gauntlett-gutowski-suryanarayana}, an explicit time-like solution was presented, in which AdS$_5$ is deformed by a 
gauge field, and two supercharges are preserved. The K\"ahler base of the five-dimensional spacetime is the Bergmann space, which is an analytic continuation 
of $\mathbb{CP}^2$. 
The five-dimensional metric takes the asymptotic form \eqref{feffer}, with boundary metric
\be\label{bdrymetricGGS}
ds^2 = - \frac{1}{\ell}\left(dt + \notsure \ell^2 \sigma_1\right)\sigma_3 + \frac 14 (\sigma_1^2 + \sigma_2^2 + \sigma_3^2 )\,,
\ee
where the $\sigma$'s are right-invariant one-forms on $S^3$:
\bea
\sigma_1 &=& \sin\phi d\theta - \sin\theta \cos\phi d\psi\,, \nonumber\\ 
\sigma_2 &=& \cos\phi d\theta + \sin\theta \sin\phi d\psi\,,  \\ 
\sigma_3 &=& d\phi +\cos\theta d\psi\ . \nonumber
\eea
This is a non-Einstein, non-conformally flat metric on $\mathbb{R} \times S^3$. In our conventions, the gauge field at the boundary reads 
\be
A = \frac{3}{2\ell}(dt+ \notsure \ell^2 \sigma_1)\,.
\label{gagge}
\ee
Here, $\notsure$ is a parameter of the solution. 
When $\notsure=0$, the gauge field is trivial, the boundary metric becomes the standard one on $\mathbb{R} \times S^3$
 (after a coordinate transformation), and the bulk spacetime is just AdS$_5$.
 
Identifying the frame as
\be
e^+ = \frac{\sigma_3}{2}\,,\qquad
e^- = -\frac{2}{\ell}(dt + \notsure \ell^2 \sigma_1) + \frac{\sigma_3}{2} \,,\qquad
w = \frac{1}{2}(\sigma_1 - i \sigma_2)\,,
\ee
we see that \eqref{bdrymetricGGS} agrees with our general description 
of section \ref{sub:coord}, with the coordinate identification
$\{y,\, u,\, x^1,\, x^2 \} \,=\, \{ -t/\ell ,\, \phi,\, \theta,\, \psi \} $. We also need to identify
\be
H = 2\,,\qquad \mathcal F = \frac 14\, , \qquad \beta = \cos \theta d\psi \,,\qquad 
\noname = -\notsure \ell \sigma_1\,,
\ee
and the metric $h_{mn}$ with the round metric on $S^2$. One can also check
 that the gauge field in (\ref{gagge})
is consistent with our general formulae in section  \ref{sub:coord}.
Using  \eqref{formulavperp}, we find that $v^\perp =  \frac 12 e^-$ and
this can be completed by choosing $e^-\cdot v = -1$, so that 
\be
v \,=\,\frac 12( e^- - e^+) \,=\, -e^0
\ee 
satisfies $d*v=0$. Finally, $a = A + \frac{3}{2} v = 0$ is consistent with (\ref{aperp}) and (\ref{genaz}). 

We checked that with these values of $v$ and $a$, the new minimal equation \eqref{eq:nm} is solved by a constant spinor 
$\epsilon_+$ satisfying the projection $\gamma^0\gamma^1 \epsilon_+ = \epsilon_+$.
This shows that the background preserves precisely 
two supercharges\footnote{Note that in order to map the frame chosen  in \cite{gauntlett-gutowski-suryanarayana} into 
the five-dimensional frame used here, one needs to perform an $r$-dependent Lorentz transformation.  
Acting on the spinors, this transforms the spinors in \cite{gauntlett-gutowski-suryanarayana}, which are independent of $r$, into 
$r$-dependent spinors, with asymptotic form given in  (\ref{eq:expansion}).
Note also that the $t$-dependence of the spinors in \cite{gauntlett-gutowski-suryanarayana} arises as a consequence of a
different gauge for $A$. In particular, in \cite{gauntlett-gutowski-suryanarayana}: $A\cdot z=0$.}.
Finally, we note that from the point of view of the boundary geometry we could deform the metric on $S^3$ in various ways. However, which deformations can be completed
to a non-singular solution in the bulk is a very hard question to address.


\subsection{Null case} 
\label{sub:null}

In the null case, $f=0$ and the bulk metric can be written as \cite{gauntlett-gutowski}
\begin{equation}\label{eq:bulknull}
ds^2 \,=\, -2\hat H^{-1}du \left(dy + \tfrac{1}{2}\mathcal F du  \right) + \hat H^2 \gamma_{mn} dx^m dx^n\,,
\end{equation}
where $\hat H$, $\gamma_{mn}$ and ${\cal F}$ depend only on $u$ and $x^m$, $m=1,2,3$, but not on $y$. Here
\begin{equation}
V = \hat H^{-1} du
\end{equation}
and by comparison with (\ref{eq:vexpr22}) we see that $\hat H^{-1} =  r^2 H^{-1} +\ldots,$ and $z= H^{-1} du$,
in agreement with the results of section \ref{sub:nontw}. 

Explicit asymptotically locally AdS solutions in the null case are also discussed in \cite{gauntlett-gutowski}. 
These are the magnetic string solutions of \cite{chamseddine-sabra,klemm-sabra}.
The boundary is $\mathbb{R}^{1,1}\times M_2$, with metric (after some obvious rescaling)
\be
ds^2 = 2\, dudy + ds^2(M_2)\ ,
\ee
and the gauge field is
\be
F = -\frac{k}{2} {\rm vol}(M_2)\ .
\ee
Here, $M_2$ is $S^2$ if $k>0$ (with radius $k^{-1/2}$), $\mathbb{T}^2$ if $k=0$, or the hyperbolic space $\mathbb{H}^2$ if $k<0$ 
(with radius $(-k)^{-1/2}$). The bulk space-time has a regular horizon when $k<0$,
while it has a naked singularity when $k>0$.
Setting  $H=1,\, \mathcal F = 0$, we find that the formulae in our section \ref{sub:nontw} are consistent with $v=0$ and $F=da$.

Notice that these bulk solutions can be easily Wick-rotated to Euclidean signature, giving boundary metrics on $\mathbb{R}^2\times M_2$.
In the case $M_2=\mathbb{H}^2$, the Wick-rotated bulk solution is non-singular, and interpolates between Euclidean 
AdS$_5$ asymptotically and $\mathbb{H}^3\times \mathbb{H}^2$ in the interior. The similar case $\mathbb{T}^2\times M_2$ was discussed in \cite{Samtleben:2012gy}.



\section{Discussion}

Motivated by the recent successful applications of localization techniques in the context of supersymmetric gauge theories on Euclidean curved manifolds, 
in this paper we have studied four-dimensional rigid supersymmetry on curved backgrounds in Lorentzian signature. We have shown that 
the backgrounds are quite different in the two cases, and in general they are not (and can not) be simply related by a Wick rotation. 
In Euclidean signature, preserved supersymmetry for a theory with an R-symmetry leads to a charged conformal Killing spinor equation, and is equivalent to the four-dimensional manifold being complex \cite{ktz,dumitrescu-festuccia-seiberg}. 
Here we demonstrated that in Lorentzian signature, solutions to the same equation are characterized by the existence of a conformal Killing vector
on the four-dimensional manifold. We have also discussed how rigid supersymmetry arises on the boundary of supersymmetric asymptotically locally
AdS solutions of minimal five-dimensional gauged supergravity, which were  analysed previously \cite{gauntlett-gutowski}. It would be very interesting to perform a similar 
comparison between Euclidean rigid supersymmetry on the boundary and Euclidean five-dimensional supergravity solutions, and we plan to address this 
in future work. Here we have illustrated the relationship between Lorentzian supersymmetry in the bulk and on the boundary in some examples
\cite{gauntlett-gutowski,lewandowski}, where the boundary metric is that of a (non-conformally flat)  $\mathbb{R} \times S^3$ or $\mathbb{R}^{1,1}\times M_2$. 
It would be interesting to construct new examples of non-singular supergravity solutions with other conformal boundary metrics and study their field theory duals.

\section*{Acknowledgments}
We wish to thank M. Porrati for interesting discussions. C.K., A.T.~and A.Z.~are supported in part by INFN, the MIUR-FIRB grant RBFR10QS5J ``String Theory and Fundamental Interactions'', and by the MIUR-PRIN contract 2009-KHZKRX. 
D.C. and D.M. acknowledge support from an  STFC grant ST/J002798/1.
D.M. is supported mainly by an EPSRC Advanced Fellowship EP/D07150X/3.

\appendix

\section{Spinor conventions}\label{SpinConventions}

In this appendix we collect our spinor conventions.
The Clifford$(1,4)$ gamma matrices $\gamma^\alpha$ satisfy
\be
\{\gamma_{\alpha},\gamma_{\beta}\} = 2\, \eta_{\alpha\beta}\,,\qquad
\gamma_{\alpha}^\dagger = \gamma_0 \gamma_\alpha \gamma_0\,,\qquad
\gamma_{\alpha}^t  = C\gamma_\alpha C^{-1}\,,
\ee
where the five-dimensional charge conjugation matrix $C$ satisfies $ C = - C^t = C^* = - C^{-1}$.
We adopt a representation of the Clifford algebra in which the first four gamma matrices are real, while $\gamma_5 = i  \gamma^0 \gamma^1 \gamma^2 \gamma^3 $ is purely imaginary. Then $\gamma^1,\gamma^2,\gamma^3$ are symmetric, while $\gamma^0$ and $\gamma^5$ are anti-symmetric. 
In this case, a consistent choice of the charge conjugation matrix is 
$C  =  i \gamma^0 \gamma^5$.
Our spinors are commuting. Furthermore, for five-dimensional spinors, the symplectic-Majorana condition is 
\be\label{symplMajo1}
\bar \epsilon^I = (\epsilon^I)^t C\,,
\ee
where $\epsilon^I$, $I=1,2$, are Dirac spinors and
we define $\bar\epsilon^I = \epsilon^{IJ} \epsilon^{J\,\dagger} \gamma^0$, with $\epsilon^{IJ}$ being the antisymmetric symbol,
such that $\epsilon^{12} = +1$. So a symplectic-Majorana pair $\epsilon^I$ carries in total eight real degrees of freedom.

\section{Intrinsic torsions and differential forms} 
\label{app:tor}

In this appendix we will explain  how to obtain the system (\ref{eq:viel}), which allows to compute the intrinsic torsions $p$ and $q$ by using differential forms and exterior differentials only, and not spinors. We will also give
explicit expressions for the differentials and covariant derivatives of the vielbein $\{z,e^-,w,\bar w\}$ 
and the two form $\omega$ corresponding to a conformal Killing spinor.
 
We start with the derivation of system (\ref{eq:viel}), consisting of  the derivatives of the elements of the vielbein. The easiest to compute is $dz$, (\ref{eq:dz}). $z$ is a spinor bilinear, as can be seen in (\ref{eq:bisp}); so its derivative can be computed in the standard way. We actually even gave its covariant derivative in (\ref{eq:nablaz}); indeed by antisymmetrizing its $\mu$ and $\nu$ indices one obtains (\ref{eq:dz}).

$dw$ and $de^-$ are trickier because $w$ and $e^-$ are not directly expressed as bilinears of $\epsilon_+$; as explained in section \ref{sub:geospinor}, they are an additional piece of data, subject to the ambiguity (\ref{eq:amb}). The two-form $\omega$, on the other hand, is a bispinor, defined in (\ref{eq:bisp}), and one can compute $d \omega$ again in a standard way; one gets (\ref{eq:domega}). Now, since $w$ is that $\omega= z\wedge w$, from $d(z\wedge w)= z \wedge d w - dz \wedge w$ one sees that 
\begin{equation}
	z \wedge dw = - 2 i {\rm Im} p \wedge z \wedge w - 2 q \wedge e^- \wedge z \ ;
\end{equation}
from this, it follows that one can write $dw$ as in (\ref{eq:dw}), for some one-form $\rho$. 

We can give an alternative characterization of $\rho$ by computing $dw$ in a different way: namely, by writing 
\begin{equation}\label{eq:wbil}
	\bar \epsilon_- \gamma_\mu e^- \epsilon_+ = -4 (e^{-})^\nu \omega_{\mu\nu}= 8 w_\mu\ 
\end{equation}
and deriving the left hand side. For this, we need to compute
\begin{align}
	D_\mu(e^- \epsilon_+) &= [D_\mu,e^-] \epsilon_+ + e^-(p_\mu \epsilon_+ + q_\mu e^- \epsilon_-)  =
	(D_\mu e^-_\nu + p_\mu e^-_\nu)\gamma^\nu \epsilon_+ = \nonumber \\
	&= -w^\mu D_\mu e^0_\nu \epsilon_- + \left(p_\mu +\frac12 z^\nu D_\mu e^-_\nu\right) e^- \epsilon_+
	\, .\label{eq:De-eps}
\end{align}
Here we have used the definition (\ref{eq:pq}) of the intrinsic torsions, and (\ref{eq:ge+}). Using this and (\ref{eq:wbil}), we get again (\ref{eq:dw}), where now we see that
\begin{equation}\label{eq:rho}
	\rho_\mu = \frac14 w^\nu D_\mu e^-_\nu \ ,\qquad
	{\rm Re} p_\mu = -\frac14 z^\nu D_\mu e^-_\nu  \ .
\end{equation}
This now suggests a way of computing $e^-$. Using 
\begin{equation}
	e^-_\mu= \frac1{16}\bar \epsilon_+ e^- \gamma_\mu e^- \epsilon_+ \ ,
\end{equation}
the expression for $D_\mu (e^- \epsilon_+)$ computed in (\ref{eq:De-eps}), and the formula for $\rho$ in (\ref{eq:rho}), we obtain (\ref{eq:de-}). 

The system of equations (\ref{eq:viel}) is general and applies to any vielbein constructed from a chiral fermion $\epsilon_+$ as explained in section \ref{sub:geospinor}; $\epsilon_+$ is not assumed to satisfy any particular differential equation. 
It is of some interest to  go on-shell and write the derivatives of the elements of a vielbein  corresponding to  a solution of the CKS equation (\ref{eq:cks}). Imposing the constraints (\ref{eq:cksconstraints}) on the torsions we have 
 \begin{align}
	\label{eq:vielcks}
	d z \,\,&=\,\,  2 {\rm Re} (q \cdot \bar w) e^-\wedge z +2 {\rm Im} (q \cdot \bar w)  i w \wedge \bar w  + 4 {\rm Re}( (q\cdot e^-) z  \wedge \bar  w)\ , \nonumber\\
	d w \,\,&=\,\,   \left( 2 i A +  \left( {\rm Re} (q\cdot \bar w) + 3 i  {\rm Im} (q\cdot \bar w)  \right) e^- - (q\cdot e^-) \bar w\right )\wedge w - 2 \sigma \wedge z  \ ,  \\
	d e^- & =\,\,  4 {\rm Re}(\sigma \wedge \bar w)  \ , \nonumber
\end{align}
where $\sigma = \rho - \frac 1 2 (q\cdot e^-) e^-$.  Equation (\ref{eq:omegacks}) easily follows from these equations. By construction, the set of equations (\ref{eq:vielcks})
implies that $z$ is conformal Killing. These equations are also interesting because they can be used to determine the gauge field $A$. 

In the new minimal case, using (\ref{eq:vp}) and the definition  $a = A + \frac 32 v$ we find
 \begin{align}
	\label{eq:vielnm}
	d z \,\,&=\,\, i\, \iota_v ( z \wedge w \wedge \bar w) , \nonumber\\
	d w \,\,&=\,\,  2 i \left(a  - \frac 3 4 (v\cdot e^-) z -  \frac 12 (v\cdot w) \bar w \right)\wedge w - 2 \sigma \wedge z  \ ,  \\
	d e^- &=\,\,  4 {\rm Re}(\sigma \wedge \bar w)  \ , \nonumber
\end{align}
from which equations (\ref{eq:dbispnm}) follows. The set of equations (\ref{eq:vielnm})
implies that $z$ is  Killing. They allow to determine uniquely the background fields $a$ and $v$. 

Finally, we also give some expressions for the covariant derivatives of the forms $z$ and $\omega$ corresponding to a solution of the CKS equation, which  have been used in the bulk to boundary comparison in section \ref{sec:bulk}.
The expressions are not particularly nice in terms of the torsions $p$ and $q$ but become simple if we replace $q$ with the vector $v$ using (\ref{eq:vp}).
This formal redefinition can be used both in the case of solutions of the CKS equation and in the case of solutions of the new minimal conditions.
As discussed in section \ref{sec:nm}, the only difference between the two cases is that, for conformal Killing spinors, $v$ has a complex part given by (\ref{eq:imv}).
We also use $a= A+\frac 32 v$.  By explicitly differentiating the bilinears $z$ and $\omega$ and using (\ref{eq:cksconstraints}), we find
\begin{align}
\label{eq:cov}
\nabla_\nu z_\mu & =  2\, {\rm Im}v_{[\mu} z_{\nu ]}  + {\rm Re}v^\tau (* z)_{\mu\nu\tau} - g_{\mu\nu} z_\tau {\rm Im} v^\tau \, , \\
\label{eq:covomega}
\nabla_\tau \omega_{\mu\nu} & = 2 i A_\tau \omega_{\mu\nu} + i (v\wedge \omega)_{\mu\nu\tau} + i (g_{\nu\tau} v^\sigma \omega_{\mu\sigma} - g_{\mu\tau} v^\sigma \omega_{\nu\sigma}) \, .
\end{align}
As expected, by symmetrizing and anti-symmetrizing and using (\ref{eq:vp}) and (\ref{eq:imv}) we recover known formulae: (\ref{eq:ckv}), the first expression in (\ref{eq:vielcks})  and 
(\ref{eq:dbispnm}).





\providecommand{\href}[2]{#2}

\end{document}